\documentclass[12pt]{iopart}

\newcommand{\PRA}{{\it Phys. Rev.} A }

\newcommand{\OC}{{\it Opt. Commun.} }
\newcommand{\JOSAB}{{\it J. Opt. Soc. Am.} B }
\newcommand{\OL}{{\it Opt. Lett. }}

\newcommand{\mycomm}[1]{}
\newcommand{\ashcomm}[1]{}
\def\COMMENTS{\renewcommand{\mycomm}[1]{\par\noindent\textcolor{red}{\emph{##1}}}
\renewcommand{\ashcomm}[1]{\par\noindent\textcolor{blue}{\emph{##1}}}}

\newcommand{\UQ}{ARC Centre of Excellence for Quantum-Atom Optics, 
School of Physical Sciences, University of Queensland, Brisbane, 
Qld 4072, Australia.}
\usepackage{graphicx}
\usepackage{bm}
\usepackage{iopams}
\usepackage[dvips]{color}

\COMMENTS

\begin{document}

\title[Tripartite entanglement and three-mode EPR correlations]{Continuous variable tripartite entanglement and Einstein-Podolsky-Rosen correlations from triple nonlinearities}

\author{M.~K. Olsen, A.~S. Bradley and M.~D. Reid}

\address{\UQ}

\begin{abstract} 
   
We compare theoretically the tripartite entanglement available from the use of three concurrent $\chi^{(2)}$ nonlinearities and three independent squeezed states mixed on beamsplitters, using the van Loock-Furusawa inequalities. We also define three-mode generalisations of the Einstein-Podolsky-Rosen paradox which are an alternative for demonstrating the inseparability of the density matrix.

\end{abstract}


\submitto{\jpb}

\ead{mko@physics.uq.edu.au}

\maketitle

\section{Introduction}
\label{subsec:intro}

Entanglement is a property which is central to quantum mechanics, with bipartite entanglement being readily producible experimentally. 
There has been some progress in the production of tripartite entangled beams, with the entanglement often obtained by mixing squeezed vacua with linear optical elements~\cite{Jing,aoki}. Other methods which create the entanglement using an actual nonlinear interaction have been proposed, using both cascaded and concurrent $\chi^{(2)}$ processes~\cite{Guo,ferraro,Nosso,ourjpb}. In this article we investigate the fundamental limits to the achievable tripartite entanglement available from both the manipulation of squeezed states with beamsplitters and from a process which utilises concurrent nonlinearities. 

A two-mode system is considered to be bipartite entangled if the system density matrix cannot be expressed as a product of density matrices for each of the two modes. The definition of tripartite entanglement for three-mode systems is a little more subtle, with different classes of entanglement having been defined, depending on how the system density matrix may be partitioned~\cite{Giedke}. The classifications range from fully inseparable, which means that the density matrix is not separable for any grouping of the modes, to fully separable, where the three modes are not entangled in any way. For the fully inseparable case, van Loock and Furusawa~\cite{vanLoock2003}, who call this genuine tripartite entanglement, have derived inequalities which are easily applicable to continuous variable processes.  In this work we will analyse two different Hamiltonian processes in terms of these inequalities, as well as in terms of three-mode Einstein-Podolsky-Rosen (EPR)~\cite{EPR} type criteria which we shall define. We note here that both these methods of detecting entanglement provide sufficient, but not necessary, conditions, so that one or the other may be more sensitive and useful in a given situation.

\section{Criteria for tripartite entanglement}
\label{sec:criteria}

We will first describe inequalities which, if violated, demonstrate that a system exhibits true continuous variable tripartite entanglement. For three modes described by the annihilation operators $\hat{a}_{j}$, where $j=1,2,3$, we define 
quadrature operators for each mode as
\begin{equation}
\hat{X}_j = \hat{a}_j+\hat{a}_j^\dag,\:\:\:
\hat{Y}_j = -i(\hat{a}_j-\hat{a}_j^\dag),
\label{eq:quaddefs}
\end{equation}
so that the Heisenberg uncertainty principle requires $V(\hat{X}_{j})V(\hat{Y}_{j})\geq 1$. 
Conditions sufficient to demonstrate continuous variable bipartite entanglement were developed by Duan \etal~\cite{Duan} and Simon~\cite{simon}. A set of conditions which are sufficient to demonstrate tripartite entanglement for any quantum state have been derived by van Loock and Furusawa~\cite{vanLoock2003}. Using our quadrature definitions, the van Loock-Furusawa conditions give a set of inequalities,
\begin{eqnarray}\label{vlf}
\eqalign{
V_{12}=V(\hat{X}_1-\hat{X}_2) + V(\hat{Y}_1+\hat{Y}_2+\hat{Y}_3) \geq& 4,\\
V_{13}=V(\hat{X}_1-\hat{X}_3) + V(\hat{Y}_1+\hat{Y}_2+\hat{Y}_3) \geq& 4,\\
V_{23}=V(\hat{X}_2-\hat{X}_3) + V(\hat{Y}_1+\hat{Y}_2+\hat{Y}_3) \geq& 4,}
\label{eq:tripart}
\end{eqnarray}
where $V(A)\equiv\langle A^2\rangle-\langle A\rangle^2$.
As shown in reference \cite{vanLoock2003}, the violation of the first condition still leaves the possibility that mode $3$ could be separated from modes $1$ and $2$, but this possibility is negated by violation of the second inequality. Starting with any one of the conditions thus shows that, if any two of these inequalities are violated, the system is fully inseparable and genuine tripartite entanglement is guaranteed. We note that genuine tripartite entanglement may still be possible when none of these inequalities is violated, due to the criteria being sufficient but not necessary.   
We also note here that the original van Loock-Furusawa correlations were written in a more complicated and general form, but for the symmetric systems we evaluate in this work, the form we have given is sufficient.

\section{Entanglement and Einstein-Podolsky-Rosen correlations}
\label{sec:EPR3}

The EPR argument was introduced in 1935 in an attempt to show that quantum mechanics not be both complete and consistent with local realism~\cite{EPR}. Schr\"odinger replied that same year by introducing the concept of entangled states which were not compatible with classical notions such as local realism~\cite{gato}. In 1989 Reid~\cite{mdr1}, and Reid and Drummond~\cite{mdr1b} proposed a physical test of the EPR paradox using optical quadrature amplitudes, which are mathematically identical to the position and momentum originally considered by EPR. Reid later expanded on this work, demonstrating that the satisfaction of the 1989 two-mode EPR criterion always implies bipartite quantum entanglement~\cite{mdr2}. It was also shown by Tan~\cite{Sze} that the existence of two orthogonal quadratures, the product of whose variances violates the limits set by the Heisenberg uncertainty principle (HUP), provides evidence of entanglement. Tan demonstrated this in the context of teleportation, with the outputs from a nondegenerate optical parameteric amplifier (OPA) mixed on a beamsplitter. In this article we extend Reid's original approach, based on an inferred HUP between two quadratures, to the case of tripartite correlations, where quadratures of three different optical modes are involved. Just as the definition of tripartite entanglement is more complex than that of bipartite entanglement, with different classes of entanglement having been defined depending on possible partitions of the system density matrix~\cite{Giedke}, we find that there is more than one way to define EPR correlations for a system exhibiting tripartite entanglement. We have used two of the possible methods in a previous publication~\cite{Nosso}, while Bowen \etal have defined a third~\cite{Bowen}. In this work we will give formal proofs that the correlations we defined previously also serve to demonstrate the presence of tripartite entanglement.

\section{Tripartite entanglement from EPR correlations}
 
{\em Genuine} tripartite entanglement is verified if we rule out any bipartition of the density matrix ($\hat{\rho}$), which is to say that the full system density matrix cannot be expressed in {\em any} of the following forms
\begin{eqnarray}
\hat{\rho}=\sum_r \hat{\rho}_r^{AB}\hat{\rho}_r^C,\hspace{1cm}\hat{\rho}=\sum_r \hat{\rho}_r^{A}\hat{\rho}_r^{BC},\hspace{1cm}\hat{\rho}=\sum_r \hat{\rho}_r^{AC}\hat{\rho}_r^B.
\end{eqnarray}
If these factorisations are ruled out then so is the fully separable form $\hat{\rho}=\sum_r \hat{\rho}_r^{A}\hat{\rho}_r^B\hat{\rho}_r^C$.
There are two forms of the criteria that we need to consider, arising from one and two mode inference. 

\subsection{Experimental two mode inference}

The first inference scheme that we will use to prove tripartite entanglement and EPR correlations involves using experimental observations of two modes to infer properties of a third. The proof for this scheme is essentially the same as for the original bipartite entanglement result of \cite{mdr2}; nevertheless, the proof exposes a freedom in the derivation that we wish to draw attention to, so we reproduce it here.
We consider the separation of the density matrix in the form 
\begin{eqnarray}\label{twomoderho}
\hat{\rho}=\sum_r \hat{\rho}_r^{A}\hat{\rho}_r^{BC}.
\end{eqnarray}
We now introduce the operators $\hat{x}^\alpha$ and $\hat{y}^\alpha$, $(\alpha\in \{A,B,C\})$ with $[\hat{x}^\alpha,\hat{y}^\alpha]=2i$. 
We see that the conditional probability of result $x^A$ for a measurement of $\hat{x}^A$ at $A$ given a simultaneous measurement of $\hat{x}^B$ at $B$ and $\hat{x}^C$ at $C$ with results $x_i^B$ and $x_i^C$ is then $P(x^A|x_i^B,x_i^C)=P(x^A,x_i^B,x_i^C)/P(x_i^B,x_i^C)$, where, assuming separability,
\begin{eqnarray}
P(x^A,x_i^B,x_i^C)=\sum_rp_rP_r(x^A)P_r(x_i^B,x_i^C).
\end{eqnarray}
Here $P_r(x^A)=\langle x^A|\hat{\rho}_r^A|x^A\rangle$, and $P_r(x_i^B,x_i^C)=\langle x_i^B,x_i^C|\hat{\rho}_r^{BC}|x_i^B,x_i^C\rangle$, where $|x^\alpha\rangle$ are eigenstates of $\hat{x}^\alpha$. Furthermore, normalisation of the density matrix requires $\sum_r p_r=1$.

We may now use the measurements of $\hat{x}^B$, $\hat{x}^C$ to infer, with some uncertainty, the value of $\hat{x}^A$. The mean of the conditional distribution is then
\begin{eqnarray}
\mu_i^{x^{A}}=\sum_{x^A}P(x^A|x_i^B,x_i^C)x^A=\sum_r \frac{p_rP_r(x_i^B,x_i^C)}{P(x_i^B,x_i^C)}\langle x^A\rangle_r,
\end{eqnarray}
where $\langle x^A\rangle_r=\sum_{x^A}P_r(x^A)x^A$. 
\par 
The variance $\Delta^2_i x^{A}$ of the distribution $P(x^A| x_i^B,x_i^C)$ is 
\begin{eqnarray}
\Delta^2_i x^{A}=\sum_r \frac{p_r P_r(x_i^B,x_i^C)}{P(x_i^B,x_i^C)}\sum_{x^A}P_r(x^A)(x^A-\mu_i^{x^A})^2.
\end{eqnarray}
Now the mean-square $\sum_{x^A}P_r(x^A)(x^A-d)^2$ is minimised by the choice $d=\langle x^A\rangle_r$, so that
\begin{eqnarray}
\Delta_i^2 x^{A}\geq\sum_r\frac{p_r P_r(x_i^B,x_i^C)}{P(x_i^B,x_i^C)}\sigma^2_r(x^{A}),
\end{eqnarray}
where $\sigma^2_r(x^{A})$ is the variance of $x^A$ over the distribution $P_r(x^A)$. Since the estimate for the result $x^{A}$ may not be optimal, we also define an error for the estimate $\Delta^2_{inf,est}\hat{x}^{A}$, so that, after averaging over the results $x_i^B$ and $x_i^C$, we find
\begin{eqnarray}
\Delta^2_{inf,est}\hat{x}^{A}\geq \Delta^2_{inf}\hat{x}^{A}\geq \sum_{x_i^B, x_i^C}P(x_i^B,x_i^C)\Delta_i^2 x^{A}=\sum_r p_r \sigma_r^2(x^{A}),
\end{eqnarray}
and similarly $\Delta^2_{inf}\hat{y}^{A}\geq \sum_r p_r \sigma_r^2(y^{A})$. Combining the two results and using the Cauchy-Schwartz inequality gives
\begin{eqnarray}
\Delta^2_{inf}\hat{x}^{A}\Delta^2_{inf}\hat{y}^{A}&\geq& \sum_r p_r \sigma_r^2(x^{A})\sum_q p_q \sigma_q^2(y^{A})\\
&\geq&\left|\sum_r p_r \sigma_r^2(x^{A})\sigma_r^2(y^{A})\right|^2.
\end{eqnarray}
For any $\hat{\rho}^{A}_r$, the uncertainty relation takes the form $\sigma_r^2(x^{A})\sigma_r^2(y^{A})\geq 1$, so that the assumed bipartition of the density matrix \eref{twomoderho} implies
\begin{eqnarray}
\Delta^2_{inf}\hat{x}^{A}\Delta^2_{inf}\hat{y}^{A}\geq 1.
\end{eqnarray}
The experimental observation of the EPR critierion $\Delta^2_{inf}\hat{x}^{A}\Delta^2_{inf}\hat{y}^{A}< 1$ rules out the bipartition \eref{twomoderho}. Tripartite entanglement is verified by ruling out all such bipartitions. The simultaneous experimental observation of the three criteria
\begin{eqnarray}
\Delta^2_{inf}\hat{x}^{A}\Delta^2_{inf}\hat{y}^{A}< 1,\\
\Delta^2_{inf}\hat{x}^{B}\Delta^2_{inf}\hat{y}^{B}< 1,\\
\Delta^2_{inf}\hat{x}^{C}\Delta^2_{inf}\hat{y}^{C}<1,
\end{eqnarray}
is sufficient to confirm tripartite entanglement. Note that the exact form of the expression involving $x_i^B$, $x_i^C$ used in the inference (in this paper we deal with expressions of the form $\hat{x}_i^B\pm \hat{x}_C$) does not enter into the derivation. We see that in fact this separability measure is entirely independent of the way information about the remaining subsystem is used to infer properties of a single mode. To be specific, if an $N$-mode system density matrix is separable in the form $\hat{\rho}=\sum_r \hat{\rho}^k_r\hat{\rho}_r^{1,\dots k-1,k+1, \dots N}$, then, regardless of the way information from $\hat{\rho}_r^{1,\dots k-1,k+1, \dots N}$ is handled the inferred variances for mode k will satisfy the uncertainty relation $\Delta^2_{inf}\hat{x}^k\Delta^2_{inf}\hat{y}^k\geq 1$, and the EPR criteria for this kind of separability follows. However, when $N>3$ there are additional forms of separability to be ruled out if genuine $N$-partite entanglement is to be confirmed. In this work we will focus solely on tripartite entanglement.

\subsection{One mode inference}

The alternative scheme uses information about one mode to infer the combined properties of the other two.
We consider the expression of the density matrix in the form 
\begin{eqnarray}\label{onemoderho}
\hat{\rho}=\sum_r \hat{\rho}_r^{AB}\hat{\rho}_r^C.
\end{eqnarray}
The conditional probability of results $x^A$ and $x^B$ for measurements of $\hat{x}^A$ and $\hat{x}^B$ at $A$ and $B$ given a simultaneous measurement of $\hat{x}^C$ at $C$ with result $x_i^C$ is $P(x^A,x^B|x_i^C)=P(x^A,x^B,x_i^C)/P(x_i^C)$, were, assuming separability
\begin{eqnarray}
P(x^A,x^B,x_i^C)=\sum_rp_rP_r(x^A,x^B)P_r(x_i^C).
\end{eqnarray}
Here $P_r(x^A,x^B)=\langle x^A,x^B|\hat{\rho}_r^{AB}|x^A,x^B\rangle$, and $P_r(x_i^C)=\langle x_i^C|\hat{\rho}_r^C|x_i^C\rangle$.
\par 
The measurements of $\hat{x}^C$ are used to infer, with some uncertainty, the combined quadrature operators $\hat{x}_{\pm }^{AB}=\hat{x}^A\pm\hat{x}^B$, $\hat{y}^{AB}_\pm=\hat{y}^A\pm\hat{y}^B$. The mean of the conditional distribution is
\begin{eqnarray}
\mu_i^{x^{AB}_\pm}=\sum_{x^A, x^B}P(x^A,x^B|x_i^C)(x^A\pm x^B)=\sum_r \frac{p_rP_r(x_i^C)}{P(x_i^C)}\langle x^A\pm x^B\rangle_r,
\end{eqnarray}
where $\langle x^A\pm x^B\rangle_r=\sum_{x^A, x^B}P_r(x^A,x^B)(x^A\pm x^B)$. 
\par 
The variance $\Delta^2_i x^{AB}_\pm$ of the distribution $P(x^A, x^B|x_i^C)$ is 
\begin{eqnarray}
\Delta^2_i x^{AB}_\pm=\sum_r \frac{p_r P_r(x_i^C)}{P(x_i^C)}\sum_{x^A, x^B}P_r(x^A,x^B)(x^A\pm x^B-\mu_i^{x^{AB}_\pm})^2.
\end{eqnarray}
Now the mean-square $\sum_{x^A, x^B}P_r(x^A,x^B)(x^A\pm x^B-d)^2$ is minimised by the choice $d=\langle x^A\pm x^B\rangle_r$, so that
\begin{eqnarray}
\Delta_i^2 x^{AB}_\pm&\geq&\sum_r\frac{p_r P_r(x_i^C)}{P(x_i^C)}\sum_{x^A, x^B}P_r(x^A,x^B)(x^A\pm x^B-\langle x^A\pm x^B\rangle_r)^2\\
&=&\sum_r\frac{p_r P_r(x_i^C)}{P(x_i^C)}\sigma^2_r(x^{AB}_\pm )
\end{eqnarray}
where $\sigma^2_r(x^{AB}_\pm)$ is the variance of $x^A\pm x^B$ over the distribution $P_r(x^A,x^B)$. Since the estimate for the results $x^{AB}_\pm$ may not be optimal, we also define an error for the estimate $\Delta^2_{inf,est}\hat{x}^{AB}_\pm$, so that, after averaging over the results $x_i^C$, we find
\begin{eqnarray}
\Delta^2_{inf,est}\hat{x}^{AB}_\pm\geq \Delta^2_{inf}\hat{x}^{AB}_\pm\geq \sum_{x_i^C}P(x_i^C)\Delta_i^2 x^{AB}_\pm=\sum_r p_r \sigma_r^2(x^{AB}_\pm),
\end{eqnarray}
and similarly $\Delta^2_{inf}\hat{y}^{AB}_\pm\geq \sum_r p_r \sigma_r^2(y^{AB}_\pm)$. Combining the two results, and using the Cauchy-Schwartz inequality gives
\begin{eqnarray}
\Delta^2_{inf}\hat{x}^{AB}_\pm\Delta^2_{inf}\hat{y}^{AB}_\pm&\geq& \sum_r p_r \sigma_r^2(x^{AB}_\pm)\sum_q p_q \sigma_q^2(y^{AB}_\pm)\\
&\geq&\left|\sum_r p_r \sigma_r^2(x^{AB}_\pm)\sigma_r^2(y^{AB}_\pm)\right|^2.
\end{eqnarray}
We now note that for any $\hat{\rho}^{AB}_r$, the uncertainty relation for the combined quadrature takes the form $\sigma_r^2(x^{AB}_\pm)\sigma_r^2(y^{AB}_\pm)\geq 4$, so that the assumed bipartition of the density matrix \eref{onemoderho} implies
\begin{eqnarray}
\Delta^2_{inf}\hat{x}^{AB}_\pm\Delta^2_{inf}\hat{y}^{AB}_\pm\geq 4.
\end{eqnarray}
The experimental observation of the EPR criterion $\Delta^2_{inf}\hat{x}^{AB}_\pm\Delta^2_{inf}\hat{y}^{AB}_\pm< 4$, as given in \eref{onemodeineq}, implies inseparability. Genuine tripartite entanglement is verified by ruling out all such bipartitions. The simultaneous experimental observation of the three criteria
\begin{eqnarray}
\Delta^2_{inf}\hat{x}^{AB}_\pm\Delta^2_{inf}\hat{y}^{AB}_\pm< 4,\\
\Delta^2_{inf}\hat{x}^{AC}_\pm\Delta^2_{inf}\hat{y}^{AC}_\pm< 4,\\
\Delta^2_{inf}\hat{x}^{BC}_\pm\Delta^2_{inf}\hat{y}^{BC}_\pm< 4,
\end{eqnarray}
is then also sufficient to establish tripartite entanglement.

\subsection{Practical criteria}

In practice one usually has access to certain moments of quadrature variables, in particular, the elements of the covariance matrix. It has been shown \cite{mdr1} that optimised linear inference based on the covariance matrix is never better than knowing the full conditional probablity distribution for modes of interest. Consequently, if EPR inequalities are violated using linear inference then the more exact criteria developed above certainly are. Hence we can find sufficent conditions which may be of great practical value, in that experiments have been designed to measure them~\cite{Ou}.

\subsubsection{Two mode inference} 

In this case we make a linear estimate $\hat{X}_{i,est}$ of the quadrature $\hat{X}_i$ for the mode $i$ from the properties of the
combined mode $j+k$, so that, for example
\begin{equation}
\hat{X}_{i,est} = a(\hat{X}_{j}\pm\hat{X}_{k})+c,
\label{eq:linearjktoi}
\end{equation}
where $a$ and $c$ are parameters which can be optimised, both experimentally and theoretically~\cite{mdr1,Ou}. It has been shown~\cite{mdr1b} that this corresponds to minimising the variance
\begin{equation}\fl
V^{inf}_{est}(\hat{X}_{i}-a(\hat{X}_{j}\pm\hat{X}_{k})) = \langle(\hat{X}_{i}-a(\hat{X}_{j}\pm\hat{X}_{k}))^{2}\rangle-\langle\hat{X}_{i}-a(\hat{X}_{j}\pm\hat{X}_{k})\rangle^{2}
\label{eq:minimisefora}
\end{equation}
with respect to $a$. The minimum is achieved when
\begin{equation}
a_{\rm min} = \frac{V(\hat{X}_{i},\hat{X}_{j}\pm\hat{X}_{k})}{V(\hat{X}_{j}\pm\hat{X}_{k})}.
\label{eq:optimuma}
\end{equation}
In the above $V(\hat{A},\hat{B})=\langle \hat{A}\hat{B}\rangle-\langle \hat{A}\rangle\langle \hat{B}\rangle$.
Defining the optimal inferred variance for $\hat{X}_i$ as $V^{inf}(\hat{X}_i)\equiv V^{inf}_{est}(\hat{X}_{i})|_{a=a_{\rm min}}$, we obtain
\begin{eqnarray}
V^{inf}(\hat{X}_{i}) &=& V(\hat{X}_{i})-\frac{\left[V(\hat{X}_{i},\hat{X}_{j}\pm \hat{X}_{k})\right]^{2}}{V(\hat{X}_{j}\pm \hat{X}_{k})}.
\end{eqnarray}
We follow the same procedure for the $\hat{Y}$ quadratures to give expressions which may be obtained by swapping each $\hat{X}$ for a $\hat{Y}$ in the above to give the optimal inferred estimate
\begin{eqnarray}
V^{inf}(\hat{Y}_{i}) &=& V(\hat{Y}_{i})-\frac{\left[V(\hat{Y}_{i},\hat{Y}_{j}\pm \hat{Y}_{k})\right]^{2}}{V(\hat{Y}_{j}\pm \hat{Y}_{k})}.
\label{eq:EPR2}
\end{eqnarray}
A demonstration of the EPR paradox can be claimed whenever it is observed that $V^{inf}_{est}(\hat{X}_{i})V^{inf}_{est}(\hat{Y}_{i})<1$, and we have shown that such an experimental outcome is possible whenever the theory predicts
\begin{equation}\label{eq:epr2}
V^{inf}(\hat{X}_{i})V^{inf}(\hat{Y}_{i}) < 1.
\label{eq:infer21}
\end{equation}
Following on from what was proven above, this demonstration for the $3$ possible values of $i$ is then sufficient to establish tripartite entanglement.

\subsection{One mode inference}

In this case one measures $V_{inf}(\hat{X}_j\pm\hat{X}_k)=V^{inf}_{est}(\hat{X}_j\pm \hat{X}_k-a\hat{X}_i)|_{a=a_{\rm min}}$. Linear inference leads to the expression for the optimal (minimum) variance in the inferred quadrature $\hat{X}_{j}+\hat{X}_{k}$
\begin{equation}
V^{inf}(\hat{X}_{j}\pm\hat{X}_{k}) = V(\hat{X}_{j}\pm\hat{X}_{k})-\frac{\left[V(\hat{X}_{i},\hat{X}_{j})\pm V(\hat{X}_{i},\hat{X}_{k})\right]^{2}}{V(\hat{X}_{i})},
\label{eq:jkinferred}
\end{equation}
which is merely a different form of the expression given in Ref.~\cite{Nosso},
\begin{equation}
V^{inf}(\hat{X}_{j}\pm \hat{X}_{k})=V(\hat{X}_{j}\pm \hat{X}_{k})-\frac{\left[V(\hat{X}_{i},\hat{X}_{j}\pm \hat{X}_{k})\right]^{2}}{V(\hat{X}_{i})}.
\label{eq:EPR1}
\end{equation}
The same expressions hold for the $Y$ quadratures, and 
it is then straightforwardly shown that the HUP requires that 
\begin{equation}
V(\hat{X}_{j}\pm \hat{X}_{k})V(\hat{Y}_{j}\pm \hat{Y}_{k})\geq 4.
\label{eq:heisenberg}
\end{equation}
There is a demonstration of this three mode form of the EPR paradox whenever $V^{inf}_{est}(\hat{X}_{j}\pm \hat{X}_{k})V^{inf}_{est}(\hat{Y}_{j}\pm \hat{Y}_{k})<4$, which is predicted to be possible when
\begin{equation}\label{onemodeineq}
V^{inf}(\hat{X}_{j}\pm \hat{X}_{k})V^{inf}(\hat{Y}_{j}\pm \hat{Y}_{k})< 4.
\label{eq:infer12}
\end{equation}
As above, this demonstration for the $3$ possible combinations also serves to establish inseparability of the density matrix.
\section{Entanglement via beamsplitters}
\label{sec:perdadetempo}
\begin{figure}[tbhp]
\begin{center} 
\includegraphics[width=0.4\columnwidth]{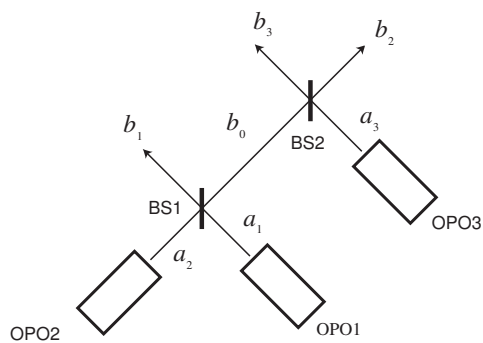}
\end{center} 
\caption{The scheme which mixes three squeezed states on two beamsplitters.}
\label{fig:mesa}
\end{figure}

It is simple to show that one quadrature squeezed state, with squeezing parameter $r$, mixed on a beamsplitter with a vacuum input, results in a bipartite entangled state with a value of $2(1+\e^{-r})$ for the Duan criterion~\cite{Duan}, where a value of less than $4$ represents bipartite entanglement. If an amplitude squeezed state is mixed with a phase-squeezed state on a beamsplitter, both with squeezing parameter $r$, this gives a value of $4\e^{-r}$ for the same criterion. In this section we will quantify one possible way in which tripartite entanglement may be obtained, using squeezed states obtained from individual $\chi^{(2)}$ processes, which are subsequently combined utilising beamsplitters.

A schematic of an apparatus which has been used by Aoki \etal~\cite{aoki} to  produce tripartite entanglement by this method is given in \fref{fig:mesa}, showing the three optical parametric oscillators (OPO) and the two beamsplitters used.
They experimentally measured continuous variable tripartite entanglement, obtaining values for the criteria of \eref{eq:tripart} which were just above $3$ using our quadrature definitions. In what follows we will first assume that OPO1 produces an ideal minimum uncertainty squeezed state of the $\hat{Y}$ quadrature and OPO2 and OPO3 produce minimum uncertainty states squeezed in their $\hat{X}$ quadratures, all with squeezing parameter $r$. With the beamsplitter BS1 having reflectivity $\mu$ and BS2 having reflectivity $\nu$, we may write expressions for the output operators $\hat{b}_{j}$ in terms of the input operators $\hat{a}_{j}$ as 
\begin{eqnarray}
\eqalign{
\hat{b}_{1} = \sqrt{1-\mu}\;\hat{a}_{1}+\sqrt{\mu}\;\hat{a}_{2},\\
\hat{b}_{2} = \sqrt{\mu(1-\nu)}\;\hat{a}_{1}-\sqrt{(1-\mu)(1-\nu)}\;\hat{a}_{2}+\sqrt{\nu}\;\hat{a}_{3},\\
\hat{b}_{3} = \sqrt{\mu\nu}\;\hat{a}_{1}-\sqrt{\nu(1-\mu)}\;\hat{a}_{2}-\sqrt{1-\nu}\;\hat{a}_{3},
}
\label{eq:bsaoki}
\end{eqnarray}
which allows us to calculate all the required output correlations in terms of the variances of the OPO outputs. The required variances are
\begin{eqnarray}
\eqalign{
V(\hat{X}_{b_{1}}) = (1-\mu)V(\hat{X}_{a_{1}})+\mu V(\hat{X}_{a_{2}}),\\
V(\hat{X}_{b_{2}}) = \mu(1-\nu)V(\hat{X}_{a_{1}})+(1-\mu)(1-\nu)V(\hat{X}_{a_{2}})+\nu V(\hat{X}_{a_{3}}),\\
V(\hat{X}_{b_{3}}) = \mu\nu V(\hat{X}_{a_{1}})+\nu(1-\mu)V(\hat{X}_{a_{2}})+(1-\nu)V(\hat{X}_{a_{3}}),\\
V(\hat{Y}_{b_{1}}) = (1-\mu)V(\hat{Y}_{a_{1}})+\mu V(\hat{Y}_{a_{2}}),\\
V(\hat{Y}_{b_{2}}) = \mu(1-\nu)V(\hat{Y}_{a_{1}})+(1-\mu)(1-\nu)V(\hat{Y}_{a_{2}})+\nu V(\hat{Y}_{a_{3}}),\\
V(\hat{Y}_{b_{3}}) = \mu\nu V(\hat{Y}_{a_{1}})+\nu(1-\mu)V(\hat{Y}_{a_{2}})+(1-\nu)V(\hat{Y}_{a_{3}}),
}
\label{eq:aokicorrelations}
\end{eqnarray}
along with the covariances 
\begin{eqnarray}
\eqalign{
V(\hat{X}_{b_{1}},\hat{X}_{b_{2}}) = \sqrt{\mu(1-\mu)(1-\nu)}\left[V(\hat{X}_{a_{1}})-V(\hat{X}_{a_{2}})\right],\\
V(\hat{X}_{b_{1}},\hat{X}_{b_{3}}) = \sqrt{\mu\nu(1-\mu)}\left[V(\hat{X}_{a_{1}})-V(\hat{X}_{a_{2}})\right],\\
V(\hat{X}_{b_{2}},\hat{X}_{b_{3}}) = \sqrt{\nu(1-\nu)}\left[\mu V(\hat{X}_{a_{1}})+(1-\mu)V(\hat{X}_{a_{2}})-V(\hat{X}_{a_{3}})\right],\\
V(\hat{Y}_{b_{1}},\hat{Y}_{b_{2}}) = \sqrt{\mu(1-\mu)(1-\nu)}\left[V(\hat{Y}_{a_{1}})-V(\hat{Y}_{a_{2}})\right],\\
V(\hat{Y}_{b_{1}},\hat{Y}_{b_{3}}) = \sqrt{\mu\nu(1-\mu)}\left[V(\hat{Y}_{a_{1}})-V(\hat{Y}_{a_{2}})\right],\\
V(\hat{Y}_{b_{2}},\hat{Y}_{b_{3}}) = \sqrt{\nu(1-\nu)}\left[\mu V(\hat{Y}_{a_{1}})+(1-\mu)V(\hat{Y}_{a_{2}})-V(\hat{Y}_{a_{3}})\right].
}
\label{eq:aokicovariances}
\end{eqnarray}  
It is straightforward to see that the modes represented by $\hat{b}_{1}$ and $\hat{b}_{0}$ can be entangled, with the Duan criterion giving
\begin{equation}
V(\hat{X}_{b_{1}}-\hat{X}_{b_{0}})+V(\hat{Y}_{b_{1}}+\hat{Y}_{b_{0}}) = 4\left[\cosh r-2\sqrt{\mu(1-\mu)}\sinh r\right].
\label{eq:aokiduan}
\end{equation}
For the case of $\mu=1/2$ this simplifies to $4\e^{-r}$, which is the well-known result for two ideal squeezed states mixed on a $50:50$ beamsplitter. In the unbalanced case we will consider here, however, $\mu=2/3$, and we find
\begin{equation}
V(\hat{X}_{b_{1}}-\hat{X}_{b_{0}})+V(\hat{Y}_{b_{1}}+\hat{Y}_{b_{0}}) = 4\cosh r - \frac{8\sqrt{2}}{3}\sinh r,
\label{eq:aokiduan01}
\end{equation}
which exhibits a maximal violation for 
\begin{equation}
r = \frac{1}{2}\log\left(\frac{1+2\sqrt{2}/3}{1-2\sqrt{2}/3}\right)\approx 1.76.
\label{eq:optr}
\end{equation}
It is of interest to note that the bipartite entanglement between these modes then disappears for $r$ a little greater than $3$, while, as we will show below, the violation of the tripartite entanglement inequalities continues to increase with $r$.

Our idealised model shows that the violation of the van Loock-Furusawa inequalities \eref{vlf} increases with the degree of squeezing, so that, for minimum uncertainty squeezed states,
\begin{eqnarray}
\fl\eqalign{
V(\hat{X}_{b_{1}}-\hat{X}_{b_{2}}) = \left[1+\mu\nu+2\sqrt{\mu(1-\mu)(1-\nu)}\right]\e^{-r}+\left[1-\mu\nu-2\sqrt{\mu(1-\mu)(1-\nu)}\right]\e^{r},\\
V(\hat{X}_{b_{1}}-\hat{X}_{b_{3}}) = \left[1+\mu-\mu\nu+2\sqrt{\mu\nu(1-\mu)}\right]\e^{-r}+\left[1-\mu+\mu\nu-2\sqrt{\mu\nu(1-\mu)}\right]\e^{r},\\
V(\hat{X}_{b_{2}}-\hat{X}_{b_{3}}) = \left[2-\mu+2\mu\sqrt{\nu(1-\nu)}\right]\e^{-r}+\mu\left[1-2\sqrt{\nu(1-\nu)}\right]\e^{r},\\
V(\hat{Y}_{b_{1}}+\hat{Y}_{b_{2}}+\hat{Y}_{b_{3}}) =\left[1+2\left(\mu\sqrt{\nu(1-\nu)}+\sqrt{\mu(1-\mu)(1-\nu)}+\sqrt{\mu\nu(1-\mu)}\right)\right]\e^{-r}\\
\hspace{3.5cm}+2\left[1-\left(\mu\sqrt{\nu(1-\nu)}+\sqrt{\mu(1-\mu)(1-\nu)}+\sqrt{\mu\nu(1-\mu)}\right)\right]\e^{r}.
}
\label{eq:idealizada}
\end{eqnarray}
With $\mu=2/3$ and $\nu=1/2$ as in Aoki \etal\cite{aoki}, we find for the van Loock-Furusawa criteria of \eref{vlf}
\begin{equation}
V_{12}=V_{13}=V_{23}=5\e^{-r},
\label{eq:finalmente}
\end{equation}
indicating clear violation of \eref{vlf} and therefore genuine tripartite entanglement for $r>0.23$.
Note that, with the quadrature definitions we use, the reported results of Ref.~\cite{aoki} are all approximately $3.2$. 

\begin{figure}
\begin{center} 
\includegraphics[width=0.8\columnwidth]{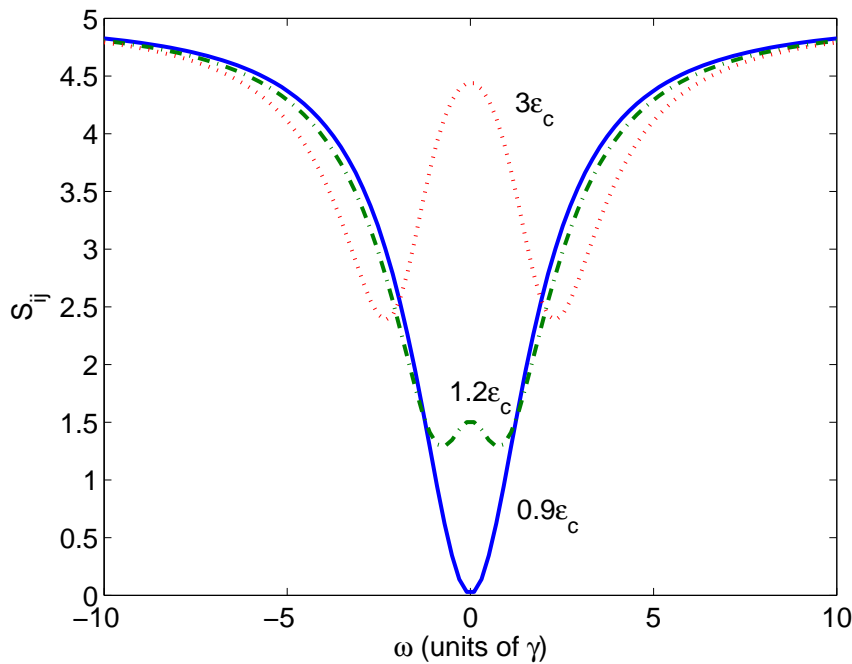}
\end{center} 
\caption{The van Loock-Furusawa spectral correlations corresponding to \eref{vlf} for the Aoki scheme, with $\mu=2/3$, $\nu=1/2$, and outputs calculated via a linearised fluctuation analysis of the standard OPO equations, with $\gamma_{a}=\gamma_{b}=1$ and $\kappa=10^{-2}$. The three curves are for different ratios of the pumping rates to the critical threshold pumping rate. All three correlations are equal for these parameters. The value $S_{ij}=4$
defines the upper boundary for true tripartite entanglement.}
\label{fig:emfrequencia}
\end{figure}

We may also now calculate the three-mode EPR correlations of \sref{sec:EPR3} for this idealised case. Again with $\mu=2/3$ and $\nu=1/2$, we find for the inference of a combined quadrature via measurements on a single quadrature
\begin{eqnarray}
V^{inf}(\hat{X}_{b_{j}}+\hat{X}_{b_{k}}) &=& \frac{6}{\mbox{e}^{r}+2\mbox{e}^{-r}},\nonumber\\
V^{inf}(\hat{Y}_{b_{j}}+\hat{Y}_{b_{k}}) &=& \frac{6}{\mbox{e}^{-r}+2\mbox{e}^{r}},
\label{eq:EPRAoki1}
\end{eqnarray}
so that
\begin{equation}
V^{inf}(\hat{X}_{b_{j}}+\hat{X}_{b_{k}})V^{inf}(\hat{Y}_{b_{j}}+\hat{Y}_{b_{k}})=\frac{36}{5+4\cosh 2r},
\label{eq:below4}
\end{equation}
which is less than $4$ for any finite value of squeezing and hence exhibits three-mode EPR correlations and entanglement. The van Loock-Furusawa inequality is not violated for $r<-\log 0.8\approx 0.22$, which is not in contradiction with the EPR criteria as both provide sufficient, but not necessary, conditions. In this particular case the EPR correlation is more sensitive to the presence of entanglement. For the inference of a single quadrature from the sum of the other two, we find
\begin{equation}
V^{inf}(\hat{X}_{i}) = \frac{3}{2\mbox{e}^{r}+\mbox{e}^{-r}},\qquad  V^{inf}(\hat{Y}_{i}) = \frac{3}{2\mbox{e}^{-r}+\mbox{e}^{r}},
\label{eq:EPEAolki2}
\end{equation}
which gives the product
\begin{equation}
V^{inf}(\hat{X}_{i})V^{inf}(\hat{Y}_{i})=\frac{9}{5+4\cosh 2r}.
\label{eq:below1}
\end{equation}
It is readily seen that this product falls below $1$ for any finite $r$ and hence gives another demonstration of EPR correlations for this system.

\begin{figure}
\begin{center} 
\includegraphics[width=0.8\columnwidth]{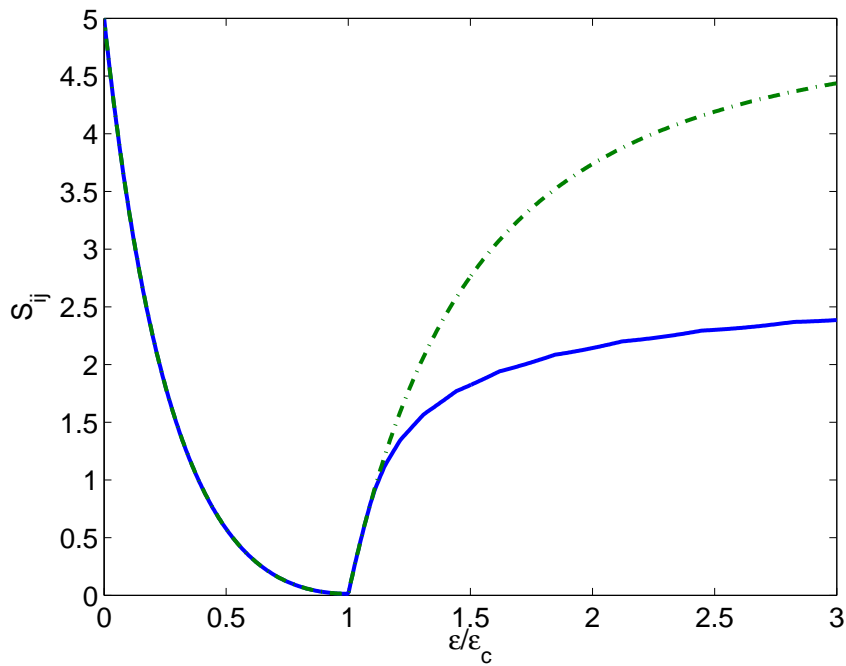}
\end{center} 
\caption{The van Loock-Furusawa correlations as for \fref{fig:emfrequencia} as a function of the ratio of the pumping rates to the critical threshold pumping rate. The solid line is the smallest value at any frequency, while the dash-dotted line is the correlation at zero frequency. All three correlations are equal for these parameters. Note that values in the immediate vicinity of $\epsilon/\epsilon_{c}=1$ are of limited value in the linearised analysis used here.}
\label{fig:omelhor}
\end{figure}

\begin{figure}
\begin{center} 
\includegraphics[width=0.8\columnwidth]{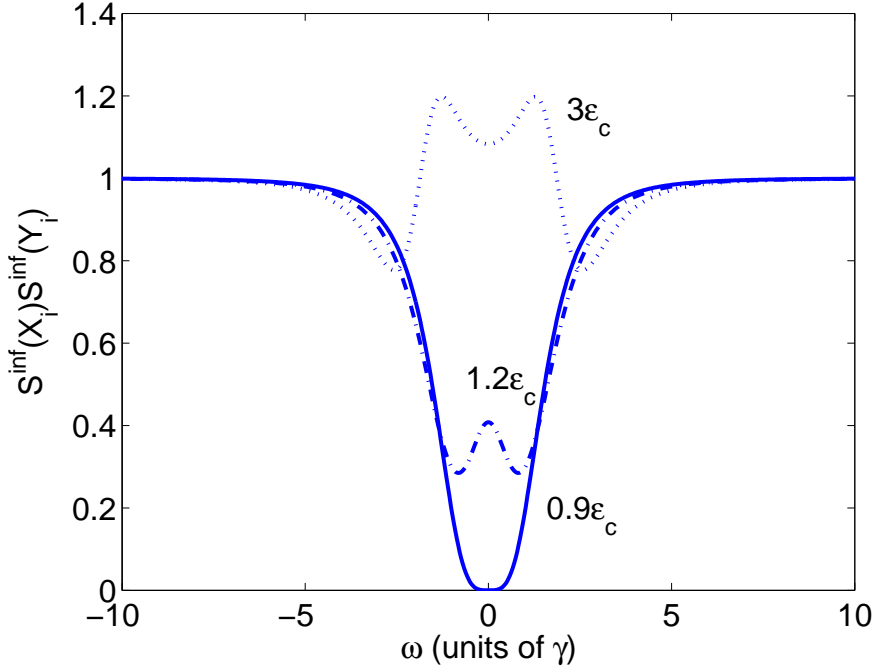}
\end{center} 
\caption{The EPR spectral correlations, $S^{inf}(\hat{X}_{i})S^{inf}(\hat{Y}_{i})$, for different ratios of the pumping rates to the critical threshold pumping rate. All parameters are the same as in \fref{fig:emfrequencia}.}
\label{fig:aokiEPR}
\end{figure}

In a more realistic analysis we must consider that OPOs do not produce minimum uncertainty squeezed states with a monotonically increasing squeezing parameter~\cite{mjc,BWO,asc}, but exhibit quite different behaviours above and below the oscillation threshold. Here we consider output quadrature amplitudes at a particular frequency shift $\omega$, so following Collett and Gardiner we define the associated spectral quadratures $\hat{X}_j(\omega)$ and $\hat{Y}_j(\omega)$, $j=1,2,3$, and the associated spectral variances. We will consider an OPO where $\hat{a}$ and $\hat{b}$ represent the signal and pump modes, with respective cavity dampings $\gamma_{a}$ and $\gamma_{b}$. With the classical pump represented by $\epsilon$ and the effective nonlinearity by $\kappa$, we find that there is a critical threshold pump value $\epsilon_{c}=\gamma_{a}\gamma_{b}/\kappa$, below which the signal mode is not macroscopically occupied. The output spectra for this OPO system are well known~\cite{mjc}, with the below threshold spectral variances for mode $\hat{a}$ being
\begin{eqnarray}\label{linspec1}
\eqalign{
S^{\rm out}(\hat{X}_{a}) = 1 + \frac{4\gamma_{a}\gamma_{b}\kappa\epsilon}{(\gamma_{a}\gamma_{b}-\kappa\epsilon)^{2}+\gamma_{b}^{2}\omega^{2}},\\
S^{\rm out}(\hat{Y}_{a}) = 1 - \frac{4\gamma_{a}\gamma_{b}\kappa\epsilon}{(\gamma_{a}\gamma_{b}+\kappa\epsilon)^{2}+\gamma_{b}^{2}\omega^{2}},
}
\label{eq:OPOabaixo}
\end{eqnarray}
while above threshold they are
\begin{eqnarray}\label{linspec2}
\eqalign{
S^{\rm out}(\hat{X}_{a}) = 1 + \frac{4\gamma_{a}^{2}(\gamma_{b}^{2}+\omega^{2})}{(2\gamma_{a}\gamma_{b}-2\kappa\epsilon+\omega^{2})^{2}+\gamma_{b}^{2}\omega^{2}},\\
S^{\rm out}(\hat{Y}_{a}) = 1 - \frac{4\gamma_{a}^{2}(\gamma_{b}^{2}+\omega^{2})}{(\omega^{2}-2\kappa\epsilon)^{2}+(2\gamma_{a}+\gamma_{b})^{2}\omega^{2}}.
\label{eq:OPOacima}
}
\end{eqnarray}
The output field $\hat{a}$ exhibits squeezing and three such outputs of three OPOs are used as the inputs $\hat{a}_1$, $\hat{a}_2$, $\hat{a}_3$ of \fref{fig:mesa}. We note here that, as the results \eref{linspec1}, \eref{linspec2} are derived using a linearised fluctuation analysis, they are not valid in the immediate region of the threshold. Given this caveat, in \fref{fig:emfrequencia} and \fref{fig:omelhor} we display the results of using these spectral variances in the expressions for the van Loock-Furusawa correlations. In these figures, the $S_{ij}(\omega)$ are the measurable output correlations which correspond to the $V_{ij}$ of \eref{eq:tripart}, so that $S_{ij}(\omega)<4$ implies genuine tripartite entanglement. It can readily be seen that the potential violation of the inequalities available from this system is much stronger than that measured so far by Aoki \etal\cite{aoki} and also that there is a large violation far above threshold. Following Reid and Drummond \cite{mdr1b}, it is also possible to use these expressions as inputs to calculate the EPR correlations of \sref{sec:EPR3} in the spectral domain. In \fref{fig:aokiEPR} we show results for the inference of one quadrature from a combination of the other two. In this case, a value of $S^{inf}(\hat{X}_i)S^{inf}(\hat{Y}_j)<1$ indicates EPR correlations, and genuine tripartite entanglement. The results for inferring the combined quadratures from the single ones are found in this case by multiplying these results by $4$, replacing, for example the $S^{inf}(\hat{X}_1)S^{inf}(\hat{Y}_1)$ with $S^{inf}(\hat{X}_2\pm\hat{X}_3)S^{inf}(\hat{Y}_2\pm\hat{Y}_3)$ and noting that the upper bound for EPR correlations is then 4. We see that this system demonstrates entanglement and the EPR paradox for a wide range of pumping strengths, these correlations persisting well into the region where the output fields are relatively intense and truly macroscopic.

\section{Entanglement via three concurrent nonlinearities}
\label{sec:verygood}

We now turn our attention to a process in which the entanglement is produced in a single nonlinear interaction which combines three concurrent nonlinearities. The Hamiltonian we will investigate in this section is derived from work 
by Pfister \etal~\cite{Pfister2004}, who raised the possibility of 
concurrent parametric down conversion in a single optical 
parametric amplifier (OPA). They also gave some solutions for equations of motion derived directly from the interaction Hamiltonian in the undepleted pump approximation, as well as experimentally observing triply coincident nonlinearities in periodically poled KTiOPO4~\cite{Pooser}.

\subsection{Properties of the Hamiltonian}
\label{subsec:viajante}

In 
this section we will also numerically solve the full equations of motion derived from the interaction Hamiltonian, as it is not possible to solve these analytically and it is known that the approximate analytic solutions and 
the full quantum solutions do not agree for arbitrary interaction strength in $\chi^{(2)}$ systems~\cite{revive,turco,shgepr}. We stress here that this analysis is not designed to give a full description of travelling-wave optical parametric amplification, which is not adequately described by our formalism (see, for example, Raymer \etal~\cite{Raymer}). However, this approach does let us examine the entanglement properties of the interaction Hamiltonian which we will later use to describe the interactions inside an optical cavity. A similar approach has been used previously to analyse, for example, the limits to squeezing and phase information in the parametric amplifier~\cite{Kinsler}.
   
\begin{figure}
\begin{center}\includegraphics[width=0.7\columnwidth]{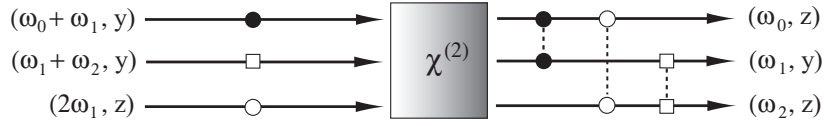}
\end{center}
\caption{Schematic of a system described by the concurrent triple 
nonlinearity interaction Hamiltonian.}
\label{fig:experiment}
\end{figure}

A schematic of the nonlinear interaction is given in 
\fref{fig:experiment}, showing the three inputs, 
which interact with the crystal to produce three output beams at 
frequencies $\omega_{0}$, $\omega_{1}$ and $\omega_{2}$, which may be 
equal. The interactions are selected to couple distinct polarisations, 
and the scheme relies on tuning the field strengths in order to 
compensate for differences in the susceptibilities since it is usually 
the case that $\chi_{yzy}\neq\chi_{zzz}$. Note that $x$ is the axis of 
propagation within the crystal. The mode described by $\hat{b}_1$ is 
pumped at frequency and polarisation $(\omega_{0}+\omega_{1},y)$ to 
produce the modes described by $\hat{a}_1$ $(\omega_{0},z)$ and 
$\hat{a}_2$ $(\omega_{1},y)$, the mode described by $\hat{b}_2$ is 
pumped at ($\omega_{1}+\omega_{2},y)$ to produce the modes described by 
$\hat{a}_2$ and $\hat{a}_3$ $(\omega_{2},z)$, while the mode described 
by $\hat{b}_3$ is pumped at $(2\omega_{1},z)$ to produce the modes 
described by $\hat{a}_{1}$ and $\hat{a}_{2}$.

The interaction Hamiltonian for the six-mode system  is then
\begin{equation}
H_{\rm int} = \rmi\hbar\left(\chi_{1}\hat{b}_1\hat{a}_1^\dag 
\hat{a}_2^\dag+\chi_{2}\hat{b}_2\hat{a}_2^\dag 
\hat{a}_3^\dag+\chi_{3}\hat{b}_3\hat{a}_1^\dag 
\hat{a}_3^\dag\right)+{\rm h.c.},
\label{eq:ham}
\end{equation}
with the $\chi_{j}$ representing the effective nonlinearities. In what 
follows, we will set $\chi_{j}=\chi$ and the high frequency input 
intensities as equal, with vacuum inputs at the lower frequencies. These 
are the conditions that we have previously found to give maximum violation of the entanglement inequalities.

\begin{figure}
\begin{center}
\includegraphics[width=0.8\columnwidth]{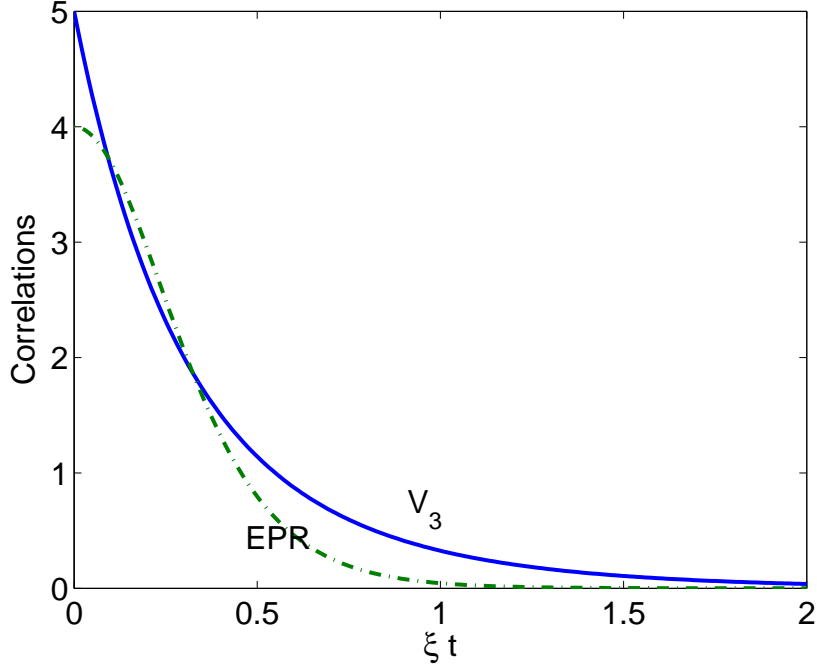}
\end{center}
\caption{Approximate analytic solutions from the Heisenberg equations of motion for the van Loock-Furusawa correlations, $V_{3}$, (any of the correlation functions of 
\eref{eq:tripart}) and the EPR correlations of \eref{eq:infer12}. The value $4$ represents the upper bound for true 
tripartite entanglement and a demonstration of the EPR paradox.}
\label{fig:heisVs}
\end{figure}

Our first step is to analytically calculate the appropriate correlations in the undepleted pumps approximation. Setting $\xi=\chi_{i}\langle\hat{b}_{i}(0)\rangle$ as a real parameter, we may solve the resulting linear Heisenberg equations to find
\begin{eqnarray}
\hat{a}_{1}(t) &=& \frac{1}{3}\left[A\hat{a}_{1}(0)+B\hat{a}_{1}^{\dag}(0)+C\hat{a}_{2}(0)+D\hat{a}_{2}^{\dag}(0)+C\hat{a}_{3}(0)+D\hat{a}_{3}^{\dag}(0)\right],\nonumber\\
\hat{a}_{1}^{\dag}(t) &=& \frac{1}{3}\left[B\hat{a}_{1}(0)+A\hat{a}_{1}^{\dag}(0)+D\hat{a}_{2}(0)+C\hat{a}_{2}^{\dag}(0)+D\hat{a}_{3}(0)+C\hat{a}_{3}^{\dag}(0)\right],
\nonumber\\ 
\hat{a}_{2}(t) &=& \frac{1}{3}\left[C\hat{a}_{1}(0)+D\hat{a}_{1}^{\dag}(0)+A\hat{a}_{2}(0)+B\hat{a}_{2}^{\dag}(0)+C\hat{a}_{3}(0)+D\hat{a}_{3}^{\dag}(0)\right],
\nonumber\\ 
\hat{a}_{2}^{\dag}(t) &=& \frac{1}{3}\left[D\hat{a}_{1}(0)+C\hat{a}_{1}^{\dag}(0)+B\hat{a}_{2}(0)+A\hat{a}_{2}^{\dag}(0)+D\hat{a}_{3}(0)+C\hat{a}_{3}^{\dag}(0)\right],
\nonumber\\ 
\hat{a}_{3}(t) &=& \frac{1}{3}\left[C\hat{a}_{1}(0)+D\hat{a}_{1}^{\dag}(0)+C\hat{a}_{2}(0)+D\hat{a}_{2}^{\dag}(0)+A\hat{a}_{3}(0)+B\hat{a}_{3}^{\dag}(0)\right],
\nonumber\\ 
\hat{a}_{3}^{\dag}(t) &=& \frac{1}{3}\left[D\hat{a}_{1}(0)+C\hat{a}_{1}^{\dag}(0)+D\hat{a}_{2}(0)+C\hat{a}_{2}^{\dag}(0)+B\hat{a}_{3}(0)+A\hat{a}_{3}^{\dag}(0)\right],
\label{eq:pfisterheisenberg}
\end{eqnarray}
where
\begin{eqnarray}
A &=& \cosh 2\xi t+2\cosh\xi t,\nonumber\\
B &=& \sinh 2\xi t-2\sinh\xi t,\nonumber\\
C &=& \cosh 2\xi t-\cosh\xi t,\nonumber\\
D &=& \sinh\xi t+\sinh 2\xi t.
\label{eq:defABCD}
\end{eqnarray}
These solutions allow us to find the quadrature variances and covariances,
\begin{eqnarray}
\eqalign{
V(\hat{X}_{a_{i}}) = \frac{1}{9}\left[(A+B)^{2}+2(C+D)^{2}\right],\nonumber\\
V(\hat{Y}_{a_{i}}) = \frac{1}{9}\left[(A-B)^{2}+2(C-D)^{2}\right],\nonumber\\
V(\hat{X}_{a_{i}},\hat{X}_{a_{j}}) = \frac{1}{9}\left(C+D\right)\left(C+D+2A+2B\right),\nonumber\\
V(\hat{Y}_{a_{i}},\hat{Y}_{a_{j}}) = \frac{1}{9}\left(C-D\right)\left(C-D+2A-2B\right).
}
\label{eq:nondepletepfister}
\end{eqnarray}
For the van Loock-Furusawa correlations, written in shorthand as $V_{3}$ since all three are equal, this then gives us
\begin{equation}
\fl
V_{3} = \frac{1}{9}\left[5(A^{2}+B^{2})+8B(D-2C)+2A(4C-8D-B)+14(C^{2}+D^{2})-20CD\right].
\label{eq:vLFheispfister}
\end{equation}
The analytic expressions for the two types of three-mode EPR correlations (one-mode and two-mode inference) are rather more complex, therefore we will present the results graphically in \fref{fig:heisVs}. We note here that each correlation is identical for any permutations of the mode indices, due to the symmetries of the Hamiltonian. The two types of EPR correlations also have exactly the same shape, with the one shown, from \eref{eq:infer12}, being identical to the correlation of \eref{eq:infer21} apart from a scaling factor of $4$. Having seen that this system is potentially a good candidate, we will now integrate a more complete version of the equations of motion, taking into account depletion of the pumping fields. 

The full equations of motion resulting from this Hamiltonian are, we gain stress, not describing an optical parametric amplifier, but contain all the dynamical information which would be described by Heisenberg equations of motion derived from the Hamiltonian of \eref{eq:ham}. 
We now map the master equation derived from \eref{eq:ham} onto a Fokker-Planck equation for the 
positive-P function~\cite{P+}, making a correspondence between the operators $\hat{a}_{i},\hat{b}_{i}$ and the classical variables $\alpha_{i},\beta_{i}$.  
We then find the appropriate stochastic 
differential equations in It\^o calculus,
\begin{eqnarray}
\eqalign{
\frac{\rmd\beta_{1}}{\rmd t} = - \chi\alpha_{1}\alpha_{2},\qquad \frac{\rmd\beta_{1}^{+}}{\rmd t} = - \chi\alpha_1^+\alpha_2^+,\nonumber\\ 
\frac{\rmd\beta_{2}}{\rmd t} = - \chi\alpha_{2}\alpha_{3},\qquad \frac{\rmd\beta_{2}^{+}}{\rmd t} = - 
\chi\alpha_{2}^{+}\alpha_{3}^{+},\nonumber\\
\frac{\rmd\beta_{3}}{\rmd t} = - \chi\alpha_{1}\alpha_{3},\qquad \frac{\rmd\beta_{3}^{+}}{\rmd t} = - 
\chi\alpha_{1}^{+}\alpha_{3}^{+},\nonumber\\ 
\frac{\rmd\alpha_{1}}{\rmd t} = \chi\left(\beta_{1}\alpha_{2}^{+} + 
\beta_{3}\alpha_{3}^{+}\right) + \sqrt{\chi\beta_1}\;\eta_{1}(t) + 
\sqrt{\chi\beta_{3}}\;\eta_3(t),\nonumber\\
\frac{\rmd\alpha_{1}^{+}}{\rmd t} = \chi\left(\beta_{1}^{+}\alpha_{2} 
+ \beta_{3}^{+}\alpha_{3}\right) + \sqrt{\chi\beta_{1}^{+}}\;\eta_{4}(t) 
+ \sqrt{\chi\beta_{3}^{+}}\;\eta_{6}(t),\nonumber\\
\frac{\rmd\alpha_{2}}{\rmd t} = \chi\left(\beta_{1}\alpha_{1}^{+} + 
\beta_{2}\alpha_{3}^{+}\right) + \sqrt{\chi\beta_2}\;\eta_{2}(t) + 
\sqrt{\chi\beta_{1}}\;\eta_{1}^{\ast}(t),\nonumber\\
\frac{\rmd\alpha_{2}^{+}}{\rmd t} = \chi\left(\beta_{1}^{+}\alpha_{1} 
+ \beta_{2}^{+}\alpha_{3}\right) + \sqrt{\chi\beta_2^{+}}\;\eta_{5}(t) + 
\sqrt{\chi\beta_{1}^{+}}\;\eta_{4}^{\ast}(t),\nonumber\\
\frac{\rmd\alpha_{3}}{\rmd t} = \chi\left(\beta_{2}\alpha_{2}^{+} + 
\beta_{3}\alpha_{1}^{+}\right) + \sqrt{\chi\beta_2}\;\eta_{2}^{\ast}(t) 
+ \sqrt{\chi\beta_{3}}\;\eta_{3}^{\ast}(t),\nonumber\\
\frac{\rmd\alpha_{3}^{+}}{\rmd t} = \chi\left(\beta_{2}^{+}\alpha_{2} 
+ \beta_{3}^{+}\alpha_{1}\right) + 
\sqrt{\chi\beta_{2}^{+}}\;\eta_{5}^{\ast}(t) + 
\sqrt{\chi\beta_{3}^{+}}\;\eta_{6}^{\ast}(t).
}
\label{eq:sde6b}
\end{eqnarray}
The complex Gaussian noise terms have the correlations
\begin{equation}
\overline{\eta_{j}} = 0, \:\:\: 
\overline{\eta_{j}^{\ast}(t)\eta_{k}(t')} = \delta_{jk}\delta(t-t').
\label{eq:Wiener}
\end{equation}
An important point is that the pairs $\alpha_{i}\;(\beta_{i})$ and $\alpha_{i}^{+}\;(\beta_{i}^{+})$ are complex conjugate only in the mean, due to the independence of the noise sources. This is necessary to allow the positive-P distribution to represent states which have a more singular distribution than the $\delta$-function distribution of a coherent state in the normal P-distribution~\cite{Crispin}. As always with the positive-P representation, averages of the variables converge to normally-ordered expectation values of the corresponding operators in the limit of a large number of stochastic trajectories.

\begin{figure}
\begin{center}
\includegraphics[width=0.8\columnwidth]{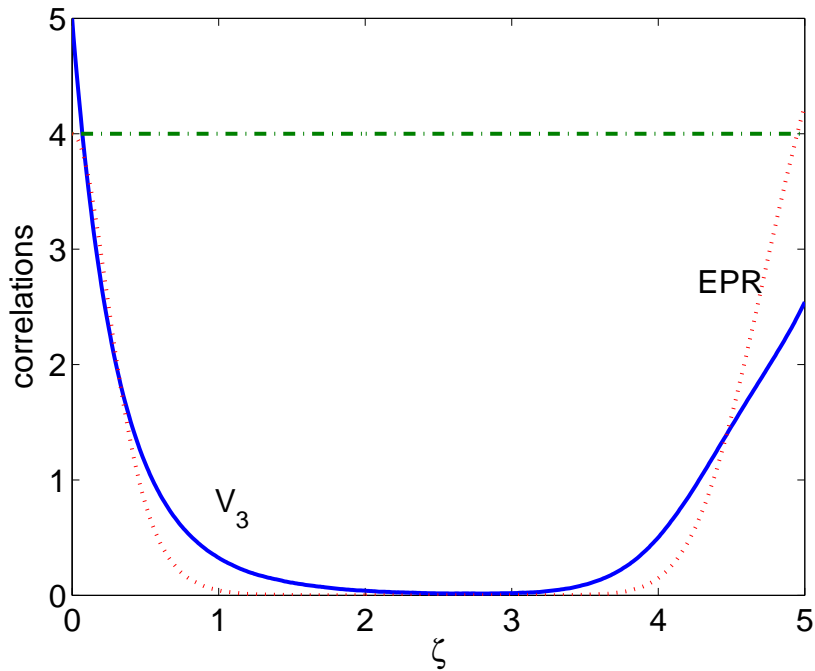}
\end{center}
\caption{Positive-P solutions from \eref{eq:sde6b} averaged over $11.92\times 10^6$ stochastic 
trajectories for the tripartite entanglement criteria and the three-mode EPR correlation. $V_{3}$ is any of the correlation functions of 
\eref{eq:tripart} and EPR is the correlation of \eref{eq:infer12}. The line at $4$ represents the upper bound for true 
tripartite entanglement.}
\label{fig:travel}
\end{figure}

The above equations were numerically integrated and averaged over 
$11.92\times 10^{6}$ trajectories, with the results for the 
correlation functions of \eref{eq:tripart} and \eref{eq:infer12} being shown in 
\fref{fig:travel}. The initial conditions were $\chi=10^{-2}$, 
$\beta_{j}(0)=10^{3}$ (real coherent states), and $\alpha_{j}(0)=0$ for $j=1,2,3$. The 
horizontal axis is a scaled interaction time, $\zeta=\chi|\beta(0)|t$. 
The minimum value of the correlations was approximately $0.02$, which is 
very close to the zero predicted by the non-depleted theory. The 
van Loock-Furusawa inequalities are strongly violated over a range of interaction length, 
although phase noise from the depleted pumps eventually means that the 
violation vanishes, as it does with the EPR correlation. The EPR correlation for inferring one mode from the combination of the other two is again equal to that shown for one-mode inference, apart from the scaling factor of $4$. For the parameters we have used, everything is symmetric, with correlations not changing under changes of the mode indices.

\subsection{Intracavity}
\label{subsec:toberejected}

\begin{figure}
\begin{center}
\includegraphics[width=0.8\columnwidth]{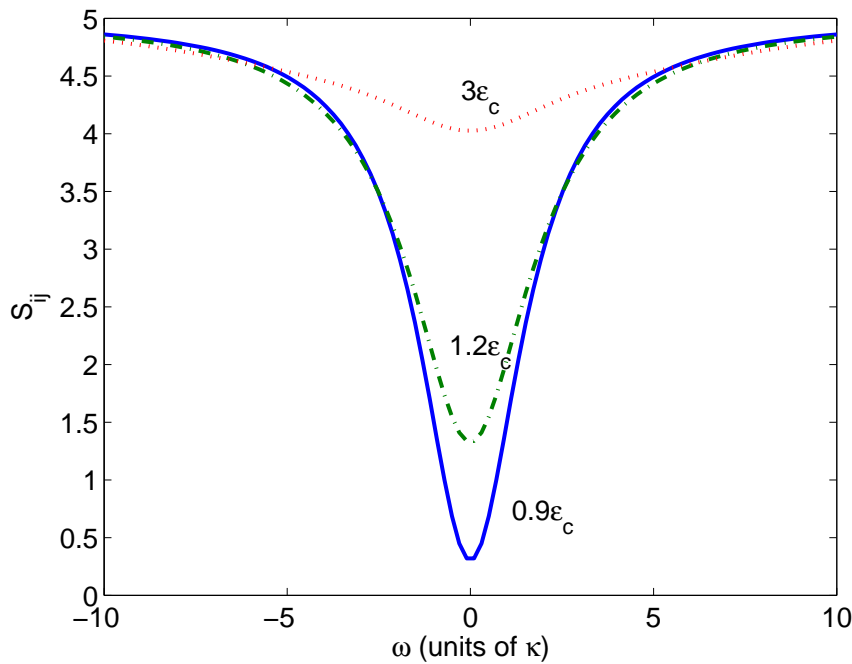}
\end{center}
\caption{The spectral correlations for different ratios of the pumping 
rate to the critical pumping rate, for the intracavity triply concurrent 
scheme. The parameter values are $\gamma=10,\:\kappa=1$ and $\chi=10^{-2}$.}
\label{fig:bombeios}
\end{figure}

\begin{figure}
\begin{center}
\includegraphics[width=0.8\columnwidth]{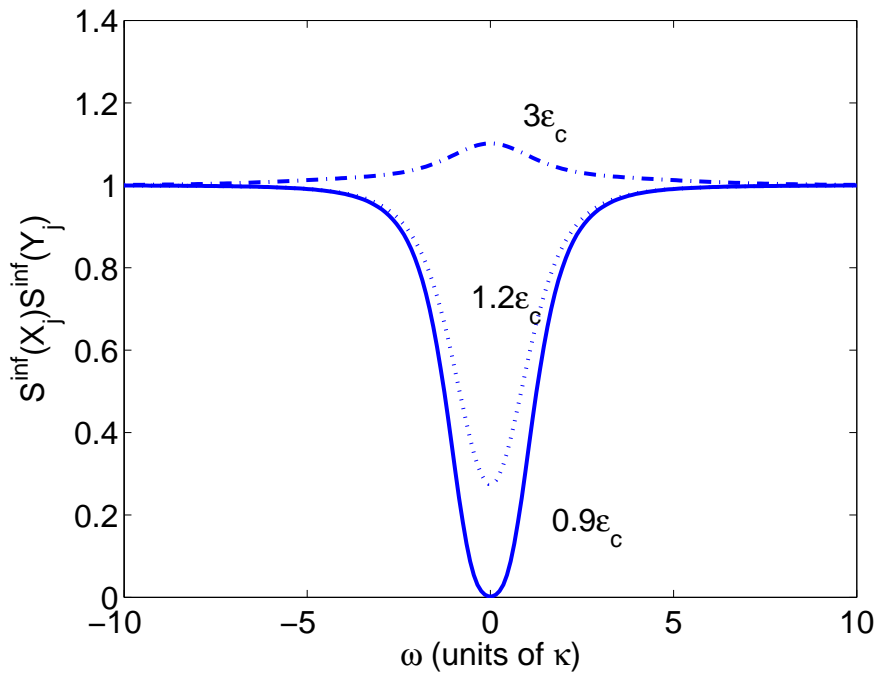}
\end{center}
\caption{The spectral EPR correlations for different ratios of the pumping 
rate to the critical pumping rate, for the intracavity triply concurrent 
scheme. The parameter values are $\gamma=10,\:\kappa=1$ and $\chi=10^{-2}$.}
\label{fig:EPRgamma}
\end{figure}

\begin{figure}
\begin{center}
\includegraphics[width=0.8\columnwidth]{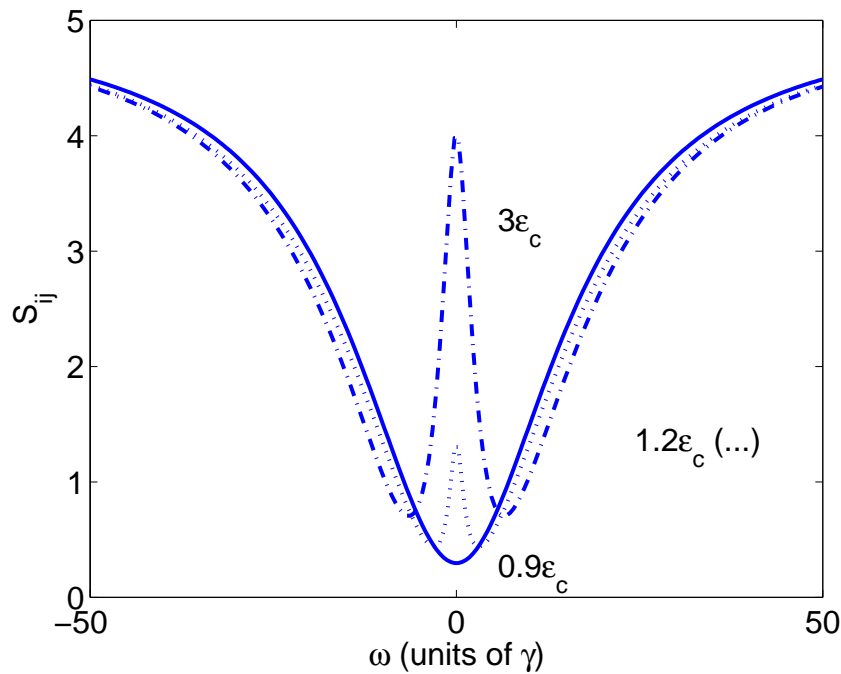}
\end{center}
\caption{The spectral correlations for different ratios of the pumping 
rate to the critical pumping rate, for the intracavity triply concurrent 
scheme. The parameter values are $\gamma=1,\:\kappa=10$ and $\chi=10^{-2}$.}
\label{fig:critkappa}
\end{figure}

\begin{figure}
\begin{center}
\includegraphics[width=0.8\columnwidth]{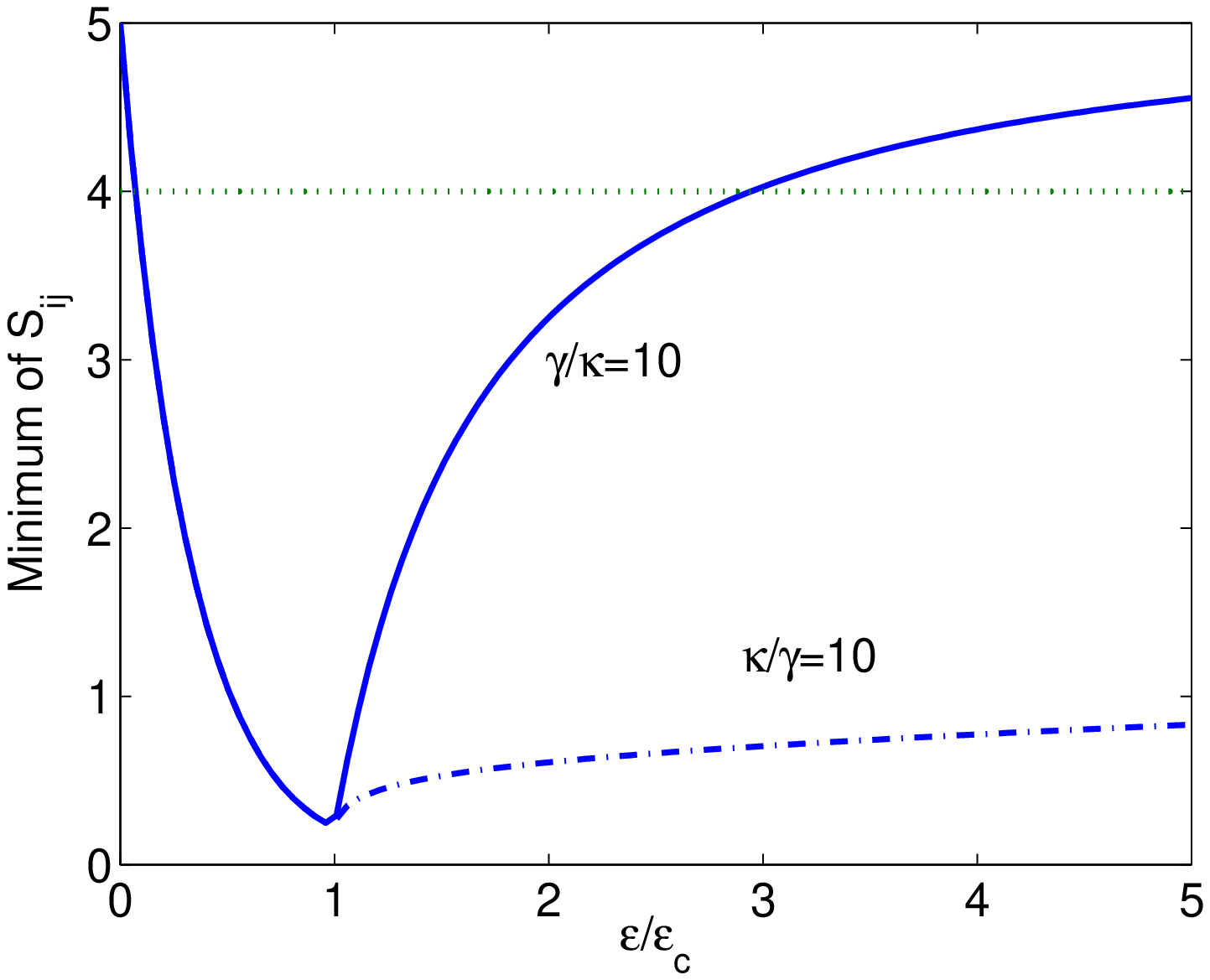}
\end{center}
\caption{The $S_{3}$ correlations as a 
function of the ratio of the cavity pumping to the threshold value and for different ratios of the cavity damping rates. The dashed line at $4$
defines the upper boundary for true tripartite entanglement. The parameter values for the solid line were 
$\gamma=10,\:\kappa=1$ and $\chi=10^{-2}$, and for the dash-dotted line, $\gamma=1,\:\kappa=10$ and $\chi=10^{-2}$.}
\label{fig:cavidade}
\end{figure}

We will now examine the experimentally more realistic case where the interaction takes place inside a pumped Fabry-Perot cavity, as
previously investigated by Bradley \etal~\cite{Nosso}. In the best 
case, where the three pumping inputs and nonlinearities are equal, 
relatively simple analytic expressions can be found for the output 
spectral correlations equivalent to \eref{eq:tripart} by following the 
usual linearised fluctuation analysis procedure. We make the proviso 
that these are not valid in the immediate region of the oscillation 
threshold, which occurs at a pump amplitude of $\epsilon^{\rm 
th}=\gamma\kappa/2\chi$, where $\gamma$ is the cavity damping rate at 
the high frequencies and $\kappa$ is the low-frequency damping rate. 
Below threshold, the steady-state solutions for the $\alpha_{j}$ are all 
zero, while $\beta_{j}^{ss}=\epsilon/\gamma$. Above threshold these 
solutions become $\beta_{j}^{ss}=\kappa/2\chi$ and $\alpha_j^{ss} = 
\sqrt{(\epsilon-\epsilon^{\rm th})/\chi}$. We note here that, due to the 
presence of the square-root, there is an ambiguity in the sign of these 
solutions. However, closer analysis shows that all must have the same 
sign, whether this is positive or negative. The full spectral 
correlations have been presented in~\cite{Nosso} for the damping ratio $\gamma/\kappa=10$ in which case the maximum violation of the van Loock-Furusawa inequalities is found at zero frequency.
These zero-frequency correlations are then found as
\begin{eqnarray}
S_{ij}^{below}(0) &=& 5 - 
\frac{8\kappa\gamma\chi\epsilon\left(4\kappa^{2}\gamma^{2}+10\kappa\gamma\chi\epsilon+7\chi^{2}\epsilon^{2}\right)}
{\left(\kappa\gamma+\chi\epsilon\right)^{2}\left(\kappa\gamma+2\chi\epsilon\right)^{2}},\nonumber\\
S_{ij}^{above}(0) &=& 
5-\frac{\kappa^{2}\gamma^{2}\left(3\kappa^{2}\gamma^{2}+6\kappa\gamma\chi\epsilon+19\chi^{2}\epsilon^{2}\right)}
{4\chi^{2}\epsilon^{2}\left(\kappa\gamma+\chi\epsilon\right)^{2}}.
\label{eq:belowandabove}
\end{eqnarray}
When the damping ratio is changed so that $\kappa/\gamma=10$, the below threshhold results are unchanged, but above threshold the spectra bifurcate so that the maximum violation is found at non-zero frequencies, as shown in \fref{fig:critkappa}. In \fref{fig:cavidade} we present the minimum of the correlation functions (maximum violation of the inequalities) for each of these cases. It is immediately obvious that the entanglement persists much further above threshhold in the bifurcated case. As this is no longer close to zero frequency, where technical noise can be a real problem, this may be a real operational advantage. 

We note here that these solutions give a limiting value of $2/9$ at 
threshold and that, even though it does not result from a valid 
analysis, this threshold value serves as an absolute minimum which can 
be expected for these correlations, in the sense that
\begin{equation}
4>S_{ij}(0)>2/9
\end{equation}
represents the region where tripartite entanglement is shown for this 
system. We also note here that, above threshold, the output modes are 
macroscopically occupied, with intensities 
$|\alpha_{j}|^{2}=(\epsilon-\epsilon^{\rm th})/\chi$, so that, especially for the case with $\kappa>\gamma$, genuine continuous variable tripartite is potentially available with intense outputs.

\begin{figure}
\begin{center}
\includegraphics[width=0.8\columnwidth]{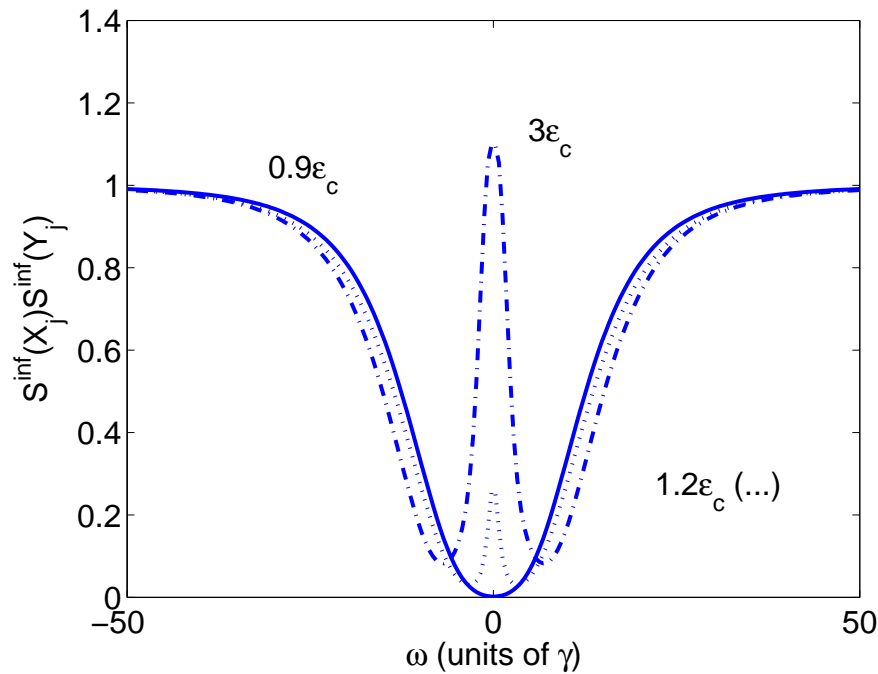}
\end{center}
\caption{The spectral EPR correlations correlations for different ratios of the pumping 
rate to the critical pumping rate, for the intracavity triply concurrent 
scheme. The parameter values are $\gamma=1,\:\kappa=10$ and $\chi=10^{-2}$.}
\label{fig:EPRkappa}
\end{figure}

\section{Conclusions}

We have examined two different interaction schemes in terms of their potential as sources of continuous variable tripartite entanglement, in terms of both the well-known van Loock-Furusawa correlations and two three-mode EPR criteria which we have developed. While both give broadly similar results, the EPR criteria, which are written in product form, may be formulated more generally than the van Loock-Furusawa criteria which generally depend on knowing the correct combinations of the quadratures involved. In the specific cases we have examined there is a symmetry which makes a simple choice possible, but this will not always be the case. 
As for the actual schemes, we have shown that the one which mixes the outputs of three OPOs on beamsplitters and that in which the three entangled modes are created in the one intracavity nonlinear material have similar performance except in the far above threshold region. Which of these two schemes is preferable for practical purposes would seem to depend more on the robustness of the experimental setup rather than any inherent advantages that either may have. On the one hand, individual OPOs are familiar technology while the type of crystal needed for the concurrent scheme is relatively new technology. On the other hand, it may prove easier to stabilise one cavity rather than having to simultaneously stabilise and synchronise three OPOs.

\ack

This research was supported by the Australian Research Council and the Queensland state government. We thank Peter Drummond for interesting discussions.

  
\end{document}